\documentclass[runningheads]{llncs}
\usepackage[T1]{fontenc}
\usepackage{graphicx,verbatim}
\usepackage{amsmath,amssymb}
\usepackage{booktabs}
\usepackage{multirow}
\usepackage{algorithm}
\usepackage{algorithmic}
\usepackage{float}
\usepackage{xurl} 
\usepackage[hidelinks]{hyperref}

\newcommand{\ours}{SCISSR}
\newcommand{\scribbleenc}{Scribble Encoder}
\newcommand{\samtwo}{SAM~2}
\newcommand{\samthree}{SAM~3}
\newcommand{\samone}{SAM}
\newcommand{\sam}{SAM}
\newcommand{\lorashort}{LoRA}

\newcommand{\lorarank}{r}

\newcommand{\dataseta}{EndoVis 2018}
\newcommand{\datasetb}{CholecSeg8k}

\begin{document}
\title{\ours{}: Scribble-Conditioned Interactive Surgical Segmentation and Refinement}
\titlerunning{\ours{}: Scribble-Conditioned Interactive Surgical Segmentation}

\author{Haonan Ping\inst{1} \and Jian Jiang\inst{1} \and Cheng Yuan\inst{1} \and Qizhen Sun\inst{1} \and Lv Wu\inst{1} \and Yutong Ban\inst{1}}
\authorrunning{H. Ping et al.}
\institute{Global College, Shanghai Jiao Tong University, Shanghai, China \\
    \email{yban@sjtu.edu.cn}}

\maketitle


\begin{abstract}
Accurate segmentation of tissues and instruments in surgical scenes is annotation-intensive due to irregular shapes, thin structures, specularities, and frequent occlusions. While \sam{} models support point, box, and mask prompts, points are often too sparse and boxes too coarse to localize such challenging targets. We present \ours{}, a scribble-promptable framework for interactive surgical scene segmentation. It introduces a lightweight \scribbleenc{} that converts freehand scribbles into dense prompt embeddings compatible with the mask decoder, enabling iterative refinement for a target object by drawing corrective strokes on error regions. Because all added modules (the \scribbleenc{}, Spatial Gated Fusion, and \lorashort{} adapters) interact with the backbone only through its standard embedding interfaces, the framework is not tied to a single model: we build on \samtwo{} in this work, yet the same components transfer to other prompt-driven segmentation architectures such as \samthree{} without structural modification. To preserve pre-trained capabilities, we train only these lightweight additions while keeping the remaining backbone frozen. Experiments on \dataseta{} demonstrate strong in-domain performance, while evaluation on the out-of-distribution \datasetb{} further confirms robustness across surgical domains. \ours{} achieves 95.41\% Dice on \dataseta{} with five interaction rounds and 96.30\% Dice on \datasetb{} with three interaction rounds, outperforming iterative point prompting on both benchmarks.

\keywords{Interactive segmentation \and Scribble annotation  \and Surgical scene segmentation}

\end{abstract}

\section{Introduction}
\label{sec:intro}

Pixel-level segmentation of surgical scenes is essential for intraoperative guidance and postoperative analysis, yet large-scale annotation is expensive due to cluttered, deformable, and overlapping anatomy and tools. Interactive segmentation reduces this cost: the annotator provides a coarse prompt, receives a predicted mask, and refines it through further interaction.

\begin{figure}[t]
    \centering
    \includegraphics[width=\linewidth]{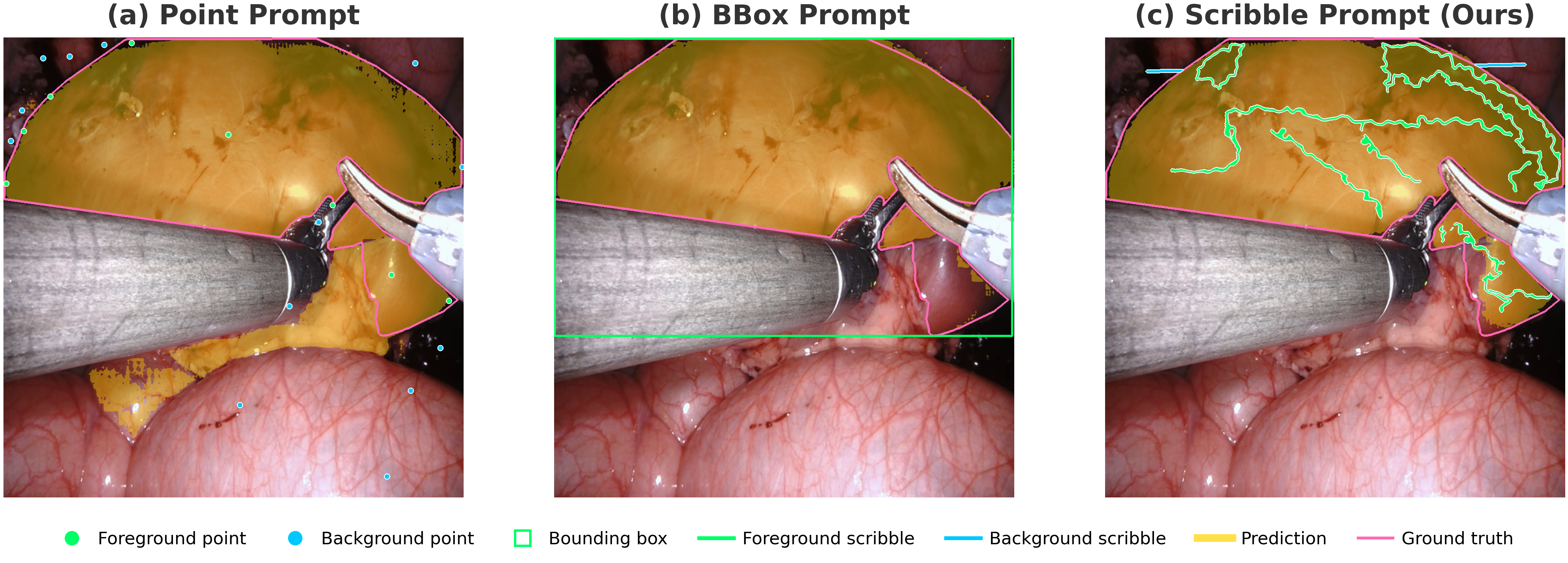}
    \caption{Motivation for scribble prompts. Points provide sparse cues and boxes enclose large background regions, while scribbles outline the target with dense spatial coverage.}
    \label{fig:motivation}
\vspace{-10pt}
\end{figure}

Foundation models such as \samone{}~\cite{sam}, \samtwo{}~\cite{sam2}, and \samthree{}~\cite{sam3} have made prompt-based segmentation practical, and medical adaptations~\cite{medsam,sammed2d} extend their reach. However, the supported prompt types (points, boxes, masks, and text) each have limitations in surgical scenes: bounding boxes cover large background areas around curved vessels or thin instruments, point clicks require many rounds for complex morphologies, and \samthree{}'s text prompts address concept-level recognition rather than instance-level delineation. As shown in Fig.~\ref{fig:motivation}, scribbles offer a natural middle ground: they trace the target with dense spatial coverage while requiring only slightly more effort than a click.


Prior interactive methods are predominantly click-based~\cite{dios,fbrs,ritm,simpleclick} (including medical variants~\cite{deepigeos,mideepseg}), and most surgical adaptations of \samone{}/\samtwo{} retain point/box prompting~\cite{medsam,sammed2d,surgicalsam,surgisam2,yuansystematic,yuanafe}. In contrast, ScribblePrompt~\cite{scribbleprompt} is primarily developed for intensity-normalized (often single-channel) biomedical images and does not leverage \samtwo{}'s memory bank, while other works use scribbles only as weak supervision~\cite{scribblesup,scribbleloss,cyclemix}.

We present \ours{}, which equips \samtwo{} with scribble-conditioned multi-round refinement for surgical scene segmentation. Our contributions are threefold: (1)~a lightweight \scribbleenc{} that maps freehand scribbles to dense prompt embeddings, enabling iterative correction via successive strokes; (2)~an architecture-agnostic design where the \scribbleenc{}, Spatial Gated Fusion, and \lorashort{}~\cite{lora} adapters attach through standard embedding interfaces, making the approach transferable to architectures such as \samthree{}; and (3)~evaluation on both \dataseta{} and the out-of-distribution \datasetb{}, confirming that scribble prompting generalizes across surgical domains (95.41\% Dice on \dataseta{}, 96.30\% on \datasetb{}).


\section{Method}
\label{sec:method}
\begin{figure}[t]
    \centering
    \includegraphics[width=\textwidth]{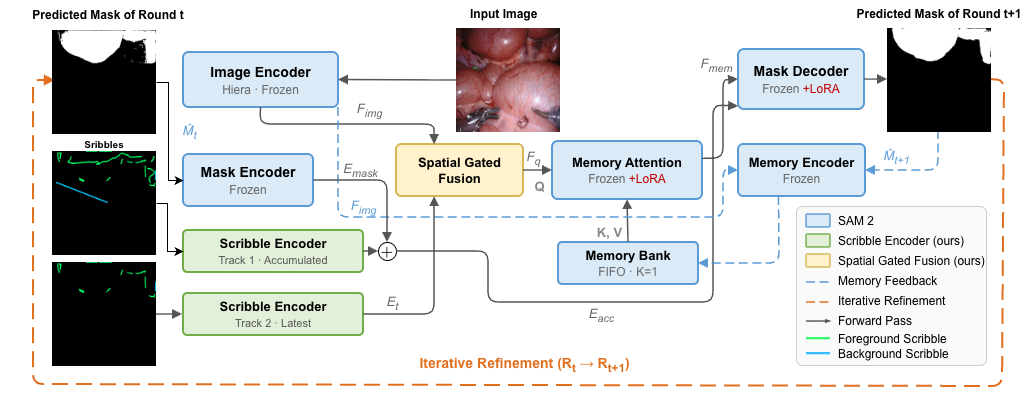}
    \caption{Overview of \ours{}. Track~1 encodes all accumulated scribbles as dense prompt embeddings for the mask decoder; Track~2 encodes only the latest correction and injects it into the Memory Attention query via Spatial Gated Fusion. The mask is iteratively refined across rounds ($R_0\rightarrow R_1\rightarrow R_2\rightarrow \cdots$). Blue: frozen \samtwo{} components; green/yellow: trainable components.}
    \label{fig:pipeline}
\vspace{-10pt}
\end{figure}

\subsection{Overview}
\label{sec:overview}
Fig.~\ref{fig:pipeline} illustrates \ours{}, a scribble-promptable framework for interactive refinement.
Although we instantiate \ours{} on \samtwo{} in this work, every added component communicates with the backbone exclusively through its standard embedding interfaces (dense prompt embeddings, memory-attention queries, and LoRA-injected projections).
Given an image $I \in \mathbb{R}^{3 \times H \times W}$ and a two-channel scribble map
$S \in \{0,1\}^{2 \times H \times W}$ (positive/foreground and negative/background),
we freeze \samtwo{}'s image encoder and introduce three lightweight components:
(i) a \scribbleenc{} that converts $S$ into dense prompt embeddings,
(ii) a Spatial Gated Fusion (SGF) module that injects the \emph{latest} correction into the memory query, and
(iii) a memory-driven iterative refinement loop that repurposes \samtwo{}'s temporal memory for multi-round correction on a single image.

A key design is a \textbf{dual-track} scribble pathway.
\textbf{Track~1 (Accumulated $\rightarrow$ Dense Prompt):} the union of scribbles across rounds is encoded and added to the mask decoder's dense prompt embedding, preserving the user's full intent.
\textbf{Track~2 (Latest $\rightarrow$ Memory Query):} only the current round's scribble is encoded and fused into the query features of Memory Attention through SGF, encouraging attention to focus on newly corrected regions.

\subsection{Scribble Encoder}
\label{sec:scribble_encoder}
The \scribbleenc{} follows \samtwo{}'s mask-prompt downscaling design.
We resize $S$ to $256 \times 256$, apply two stride-2 convolution blocks
($2{\times}2$ Conv $\rightarrow$ LayerNorm2d $\rightarrow$ GELU), and a final $1{\times}1$ projection to $D{=}256$,
producing a dense embedding $E_S \in \mathbb{R}^{256 \times 64 \times 64}$ aligned with \samtwo{}'s image-embedding resolution.
We zero the embedding when the channel input is empty.

\subsection{Spatial Gated Fusion}
\label{sec:sgf}

The Spatial Gated Fusion (SGF) module injects the latest scribble into the query features of Memory Attention. A binary hard gate $g_h \in \{0,1\}$ disables the fusion branch when no scribble is provided. When active, image features $F \in \mathbb{R}^{D \times H' \times W'}$ and scribble embedding $E_S$ are concatenated and processed by a spatial mixing operator:
\[
\texttt{SpatialMix}(\texttt{Concat}(F,\,E_S))
= \phi_{\text{dw}} \Big(\phi_{1\times1}(\texttt{Concat}(F,\,E_S))\Big),
\]
where $\phi_{1\times1}$ reduces $2D$ channels back to $D$ (with GroupNorm and GELU), and $\phi_{\text{dw}}$ is a $7{\times}7$ depthwise-separable convolution block that propagates the scribble signal spatially. A learnable scalar $\alpha$ (initialized to zero) controls the fusion strength:
\begin{equation}
F' = F + \alpha \cdot g_h \cdot \texttt{SpatialMix}\!\left(\texttt{Concat}\!\left(F,\, g_h \cdot E_S\right)\right).
\end{equation}
Because $\alpha$ is initialized to zero, SGF acts as an identity mapping at the start of training and leaves the pretrained features unchanged.

\subsection{Memory-Driven Iterative Refinement}
\label{sec:iterative}

\ours{} repurposes \samtwo{}'s temporal memory for multi-round correction on a single image. Image features are computed once and cached. Track~1 accumulates all scribbles (channel-wise maximum) as a dense prompt, while Track~2 injects only the latest correction via SGF. The Memory Bank retains only the previous round's encoded features as a single memory entry for cross-attention. This suffices because Track~1 already preserves the full interaction history. The complete pseudocode is given in Algorithm~\ref{alg:iterative} (Appendix~\ref{app:algorithm}).

\subsection{Toggleable \lorashort{} for Scribble Adaptation}
\label{sec:lora}

We keep \samtwo{}'s image encoder entirely frozen. \lorashort{} adapters~\cite{lora} are inserted into the query and value projections of all multi-head attention layers in the \textbf{mask decoder} and \textbf{memory attention} module, with $B$ zero-initialized so that $\Delta W{=}0$ at the start of training. 
Importantly, these adapters are toggleable: we enable them for scribble-conditioned refinement and can disable them for standard video propagation by setting the LoRA scaling to zero.

\subsection{Training Pipeline}
\label{sec:training}

Training proceeds in two stages. \textbf{Stage~1} trains the \scribbleenc{} and mask decoder \lorashort{} without memory or SGF, using multi-round ($T{=}3$) scribble-to-mask prediction with oracle corrections. \textbf{Stage~2} jointly trains all four components (Scribble Encoder, SGF, mask decoder \lorashort{}, Memory Attention \lorashort{}) with the full iterative pipeline. For each training sample, we unroll $T{=}3$ rounds: initial scribbles at $t{=}0$ and error-driven corrective scribbles for $t{>}0$ (Sec.~\ref{sec:scribble_gen}).

The loss at each round combines Focal Loss and Dice Loss with linearly increasing round weights $w_t = t + 1$:
\begin{equation}
    \mathcal{L}_{\text{total}} = \sum_{t=0}^{T-1} \frac{w_t}{\sum_j w_j} \Big[ 20 \cdot \mathcal{L}_{\text{focal}}(\hat{M}_t, M^*) + \mathcal{L}_{\text{dice}}(\hat{M}_t, M^*) \Big].
\end{equation}

\subsection{Scribble Generation}
\label{sec:scribble_gen}

Since large-scale scribble annotations are unavailable, we synthesize scribbles from ground-truth masks. An Adaptive Scribble Generator selects one of four types per connected component based on geometric cues: \emph{centerline} (skeleton strokes), \emph{wave skeleton} (oscillating centerline), \emph{contour} (boundary-following with inward offset), and \emph{line} (negative strokes for false positives), with mild spatial perturbations to mimic freehand variability. For correction rounds, false negatives receive positive geometry-aware scribbles and false positives receive negative cross-out strokes~\cite{scribbleprompt}.

\section{Experiments}
\label{sec:experiments}

\subsection{Datasets and Metrics}
\label{sec:datasets}

We evaluate on two laparoscopic surgical datasets.
\textbf{EndoVis18}~\cite{endovis2018} provides 15~training and 4~test sequences of robotic nephrectomy (2{,}235 / 999 frames, 10~foreground classes, 4{,}616 test samples). As the model is trained on this dataset, it serves as the in-distribution (ID) benchmark.
\textbf{CholecSeg8k}~\cite{cholecseg8k} provides 8{,}080 cholecystectomy frames with 12~foreground classes; we hold out 1{,}679 frames (8{,}800 test samples). No CholecSeg8k data is seen during training, making it an out-of-distribution (OOD) benchmark.
We report IoU and Dice (\%) per round as \textbf{sample-average} (\emph{mIoU/mDice}) and \textbf{class-average} (\emph{cIoU/cDice}); $R{k}$ denotes the prediction after $k$ refinement rounds.

\subsection{Implementation Details}
\label{sec:implementation}

We build on \samtwo{} Tiny as the backbone. \lorashort{} adapters with rank $\lorarank = 8$ and scaling factor $\alpha / \lorarank = 2$ are inserted into the query and value projections of all multi-head attention layers in the mask decoder and Memory Attention module; the image encoder remains entirely frozen. We train for 10 epochs using the AdamW optimizer with a weight decay of 0.01, a batch size of 2, and automatic mixed-precision training (AMP) on a single NVIDIA RTX 4090 (24GB), with learning rates of $1{\times}10^{-4}$ for Stage-2 newly added modules and $1{\times}10^{-5}$ for Stage~1 modules.

\subsection{Comparison with Baselines}
\label{sec:comparison}

We compare against point-based iterative methods (\samtwo{} Tiny and \samthree{}, each with 1\,pt/CC and 10\,pt/ch protocols), single bounding box baselines (\samtwo{} Tiny, \samthree{}, and MedSAM2~\cite{Ma2025MedSAM2SA}), all evaluated under the same fixed-round automated protocol (Sec.~\ref{sec:eval_protocol}). 1\,pt/CC uses one click per error-region connected component (at its centroid), whereas 10\,pt/ch provides up to 10 positive and 10 negative clicks per round (prioritizing connected-component centroids when selecting points).

\begin{table}[t]
\centering
\caption{Comparison on \dataseta{} (ID, $N{=}4{,}616$) and \datasetb{} (OOD, $N{=}8{,}800$). mIoU / mDice (\%). $R{k}$: after $k$ refinement rounds. N/A: not applicable.}
\label{tab:main_allrounds}
\setlength{\tabcolsep}{2.5pt}
\scriptsize
\resizebox{\linewidth}{!}{%
\begin{tabular}{l cc cc cc cc cc}
\toprule
\multirow{2}{*}{Method}
& \multicolumn{6}{c}{\dataseta{} (ID)}
& \multicolumn{4}{c}{\datasetb{} (OOD)} \\
\cmidrule(lr){2-7}\cmidrule(lr){8-11}
& \multicolumn{2}{c}{\textbf{R0}} & \multicolumn{2}{c}{\textbf{R2}} & \multicolumn{2}{c}{\textbf{R4}}
& \multicolumn{2}{c}{\textbf{R0}} & \multicolumn{2}{c}{\textbf{R2}} \\
\cmidrule(lr){2-3}\cmidrule(lr){4-5}\cmidrule(lr){6-7}\cmidrule(lr){8-9}\cmidrule(lr){10-11}
& mIoU & mDice & mIoU & mDice & mIoU & mDice & mIoU & mDice & mIoU & mDice \\
\midrule
\multicolumn{11}{l}{\textit{Point Prompt Baselines}} \\
SAM2 Tiny (1pt/CC)        & 56.47 & 68.31 & 62.73 & 73.77 & 55.83 & 66.73 & 62.18 & 72.30 & 62.18 & 73.44 \\
SAM2 Tiny (10pt/ch)       & 62.09 & 73.93 & 71.17 & 80.39 & 63.15 & 72.08 & 67.65 & 77.84 & 75.02 & 83.18 \\
SAM3 (1pt/CC)             & 57.29 & 69.18 & 59.70 & 70.36 & 51.32 & 60.24 & 56.74 & 67.24 & 52.62 & 61.69 \\
SAM3 (10pt/ch)            & 58.69 & 71.09 & 62.52 & 72.59 & 40.38 & 47.95 & 57.36 & 69.23 & 68.86 & 78.69 \\
\midrule
\multicolumn{11}{l}{\textit{Bounding Box Baselines}} \\
\samtwo{} Tiny (BBox)     & 64.11 & 73.23 & \multicolumn{2}{c}{N/A} & \multicolumn{2}{c}{N/A} & 76.22 & 84.62 & \multicolumn{2}{c}{N/A} \\
\samthree{} (BBox)        & 69.13 & 77.97 & \multicolumn{2}{c}{N/A} & \multicolumn{2}{c}{N/A} & 77.85 & 85.78 & \multicolumn{2}{c}{N/A} \\
MedSAM2 (BBox)            & 45.87 & 54.54 & \multicolumn{2}{c}{N/A} & \multicolumn{2}{c}{N/A} & 76.74 & 82.69 & \multicolumn{2}{c}{N/A} \\
\midrule
\multicolumn{11}{l}{\textit{Supervised Baselines (trained on \dataseta{})}} \\
U-Net~\cite{unet}         & 50.70 & 61.50 & \multicolumn{2}{c}{N/A} & \multicolumn{2}{c}{N/A} & \multicolumn{2}{c}{N/A} & \multicolumn{2}{c}{N/A} \\
UPerNet~\cite{upernet}    & 58.40 & 66.80 & \multicolumn{2}{c}{N/A} & \multicolumn{2}{c}{N/A} & \multicolumn{2}{c}{N/A} & \multicolumn{2}{c}{N/A} \\
HRNet~\cite{hrnet}        & 63.30 & 71.80 & \multicolumn{2}{c}{N/A} & \multicolumn{2}{c}{N/A} & \multicolumn{2}{c}{N/A} & \multicolumn{2}{c}{N/A} \\
SegFormer~\cite{segformer} & 63.00 & 71.90 & \multicolumn{2}{c}{N/A} & \multicolumn{2}{c}{N/A} & \multicolumn{2}{c}{N/A} & \multicolumn{2}{c}{N/A} \\
SegNeXt~\cite{segnext}    & 64.30 & 72.50 & \multicolumn{2}{c}{N/A} & \multicolumn{2}{c}{N/A} & \multicolumn{2}{c}{N/A} & \multicolumn{2}{c}{N/A} \\
STSwin-CL~\cite{stswincl} & 63.60 & 72.00 & \multicolumn{2}{c}{N/A} & \multicolumn{2}{c}{N/A} & \multicolumn{2}{c}{N/A} & \multicolumn{2}{c}{N/A} \\
LSKA-Net~\cite{lskanet}   & 66.20 & 75.30 & \multicolumn{2}{c}{N/A} & \multicolumn{2}{c}{N/A} & \multicolumn{2}{c}{N/A} & \multicolumn{2}{c}{N/A} \\
TAFPNet~\cite{yuanafe}    & 82.60 & 89.90 & \multicolumn{2}{c}{N/A} & \multicolumn{2}{c}{N/A} & \multicolumn{2}{c}{N/A} & \multicolumn{2}{c}{N/A} \\
\midrule
\multicolumn{11}{l}{\textit{Ours (Scribble Prompt)}} \\
\ours{} (Adaptive)        & 75.72 & 85.18 & 88.32 & 93.49 & 90.18 & 94.61 & 83.42 & 90.24 & 92.30 & 95.82 \\
\ours{} (Contour)         & 79.42 & 87.63 & 90.03 & 94.51 & \textbf{91.60} & \textbf{95.41} & 84.25 & 90.75 & \textbf{93.22} & \textbf{96.30} \\
\ours{} (Wave)            & 73.55 & 83.66 & 88.69 & 93.72 & 90.36 & 94.39 & 82.49 & 89.57 & 92.04 & 95.67 \\
\ours{} (Centerline)      & 74.16 & 84.13 & 85.31 & 91.58 & 86.55 & 92.30 & 83.00 & 90.00 & 90.80 & 94.88 \\
\bottomrule
\end{tabular}%
}
\vspace{-10pt}
\end{table}

Table~\ref{tab:main_allrounds} reports results on both datasets. On \dataseta{}, the contour model reaches 91.60\% mIoU at R4, while point-based methods often degrade with more rounds, with R4 metrics generally lower than R2. Bounding box and supervised baselines do not support iterative refinement and are therefore marked N/A for later rounds. On \datasetb{} (OOD), our adaptive model reaches 92.30\% mIoU at R2 without any CholecSeg8k training data, exceeding box baselines (76--78\% mIoU) by over 14\,pp. Although our model is also trained exclusively on \dataseta{}, it generalizes well to the unseen \datasetb{} domain, whereas supervised baselines trained on the same data are not directly applicable to \datasetb{} due to differences in the label set. Class-average breakdowns and additional OOD tables are provided in Appendix~\ref{app:additional_tables}.

\noindent\textbf{Convergence efficiency.}
For interactive annotation, the practical value of a method depends on how quickly it reaches acceptable quality. Table~\ref{tab:convergence} summarizes success rates and mean rounds on \dataseta{}, showing that \ours{} consistently converges faster and succeeds more often than point-based baselines across Dice thresholds.

\begin{table}[t]
\centering
\caption{Convergence efficiency on \dataseta{} ($N{=}4{,}616$). Success: \% of masks reaching the Dice threshold within 4 rounds.}
\label{tab:convergence}
\setlength{\tabcolsep}{4pt}
\scriptsize
\begin{tabular}{l l cc ccccc}
\toprule
\multirow{2}{*}{Method} & \multirow{2}{*}{Dice $\geq$} & \multirow{2}{*}{Success} & \multirow{2}{*}{Mean Rnd} & \multicolumn{5}{c}{Cumulative \% by round} \\
\cmidrule(lr){5-9}
& & (\%) & & \textbf{R0} & \textbf{R1} & \textbf{R2} & \textbf{R3} & \textbf{R4} \\
\midrule
\multirow{3}{*}{SAM2 Tiny (1pt/CC)}
 & 0.75 & 75.5 & 1.51 & 50.3 & 66.4 & 72.2 & 74.5 & 75.5 \\
 & 0.85 & 56.4 & 1.76 & 31.2 & 45.3 & 51.6 & 54.7 & 56.4 \\
 & 0.90 & 40.9 & 1.89 & 19.5 & 31.2 & 36.9 & 39.4 & 40.9 \\
\midrule
\multirow{3}{*}{SAM3 (1pt/CC)}
 & 0.75 & 71.6 & 1.44 & 51.9 & 64.2 & 68.4 & 70.4 & 71.6 \\
 & 0.85 & 52.9 & 1.75 & 30.1 & 42.2 & 48.4 & 51.2 & 52.9 \\
 & 0.90 & 39.7 & 2.04 & 17.6 & 27.8 & 34.2 & 37.8 & 39.7 \\
\midrule
\multirow{3}{*}{\ours{} (Adaptive)}
 & 0.75 & \textbf{99.0} & \textbf{1.37} & 70.2 & 92.9 & 97.3 & 98.5 & 99.0 \\
 & 0.85 & \textbf{94.7} & \textbf{1.68} & 48.4 & 81.5 & 90.7 & 93.4 & 94.7 \\
 & 0.90 & \textbf{87.7} & \textbf{1.98} & 33.3 & 66.9 & 79.2 & 85.2 & 87.7 \\
\bottomrule
\end{tabular}
\end{table}

\begin{figure}[t]
    \centering
    \includegraphics[width=\linewidth]{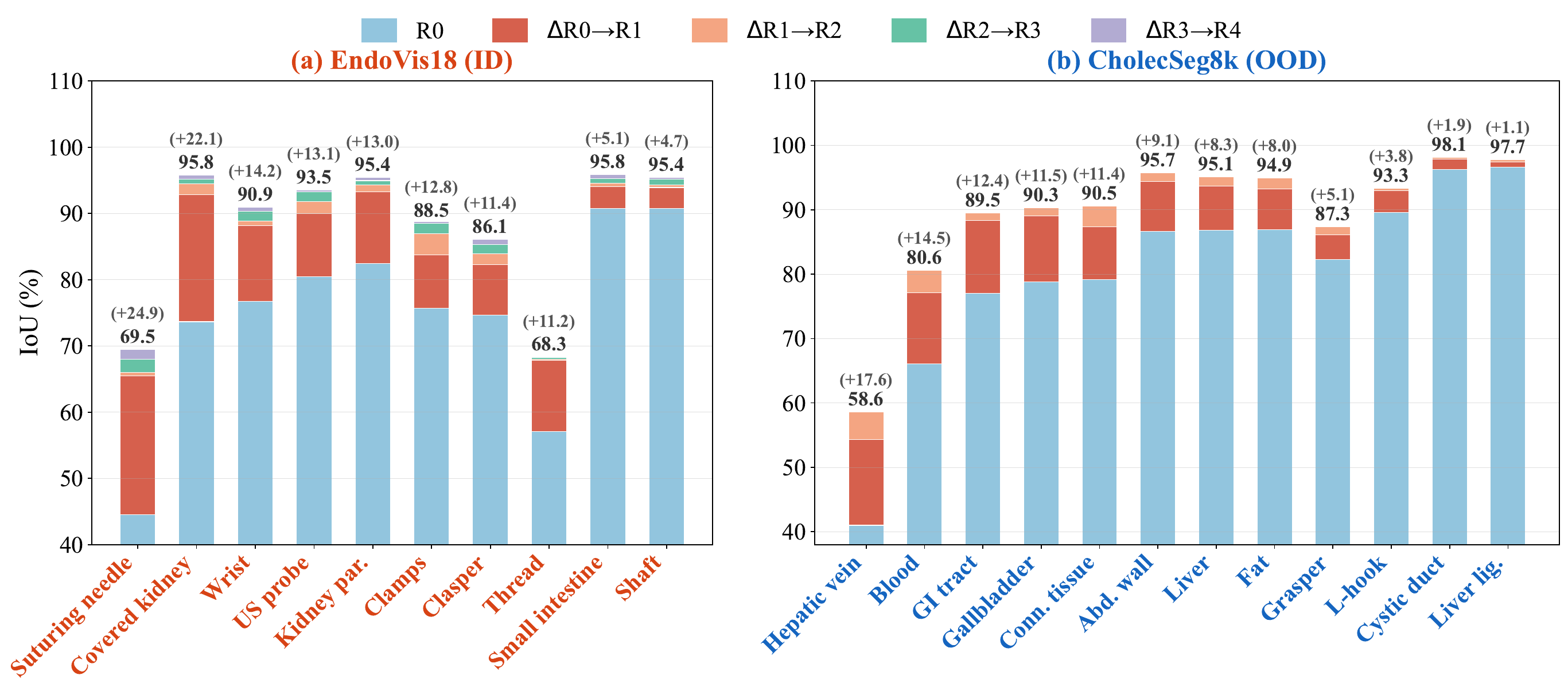}
    \caption{Per-class IoU with incremental refinement gains. (a)~\dataseta{} (contour, R0$\to$R4). (b)~\datasetb{} (OOD, R0$\to$R2). Numbers at bar tops: final IoU and total gain.}
    \label{fig:per_class}
\vspace{-10pt}
\end{figure}

\noindent\textbf{Per-class analysis.}
Fig.~\ref{fig:per_class} shows per-class results. On \dataseta{}, classes with irregular geometry benefit most: \emph{suturing needle} ($+$24.92\,pp), \emph{covered kidney} ($+$22.15\,pp), and \emph{wrist} ($+$14.16\,pp). On \datasetb{} (OOD), the model generalizes across all 12 classes without any CholecSeg8k training data. Full per-class tables for all strategies and baselines are in Appendix~\ref{app:per_class}.

\subsection{Automated Evaluation Protocol}
\label{sec:eval_protocol}

We use an automated protocol to simulate iterative interaction. An initial scribble is generated from the ground-truth mask (Sec.~\ref{sec:scribble_gen}); the model predicts a mask and, if IoU $< \tau$, corrective scribbles are generated on error regions for the next round, repeating for at most $T$ rounds. Point-click baselines follow the same protocol with clicks at the center of each error-region connected component.

\noindent\textbf{Limitations of point prompts.}
One might attribute our gains to scribbles containing more labeled pixels. However, for point prompting, both 1pt/CC and the denser 10pt/ch protocol in Table~\ref{tab:main_allrounds} can degrade as rounds increase, despite receiving more clicks. This indicates that point quantity alone is insufficient; the structured spatial layout of scribbles provides richer shape and boundary cues than isolated clicks (see Appendix~\ref{app:point_density} for a detailed density analysis).







\subsection{Ablation Studies}
\label{sec:ablation}

\noindent\textbf{Scribble Generation strategy.}
We evaluate four strategies (Sec.~\ref{sec:scribble_gen}) on \dataseta{} and \datasetb{}.
Table~\ref{tab:main_allrounds} shows that contour-only performs best along all rounds, while centerline-only strokes lag, indicating boundary cues more effective for refinement.

\begin{figure}[t]
    \centering
    \includegraphics[width=\linewidth]{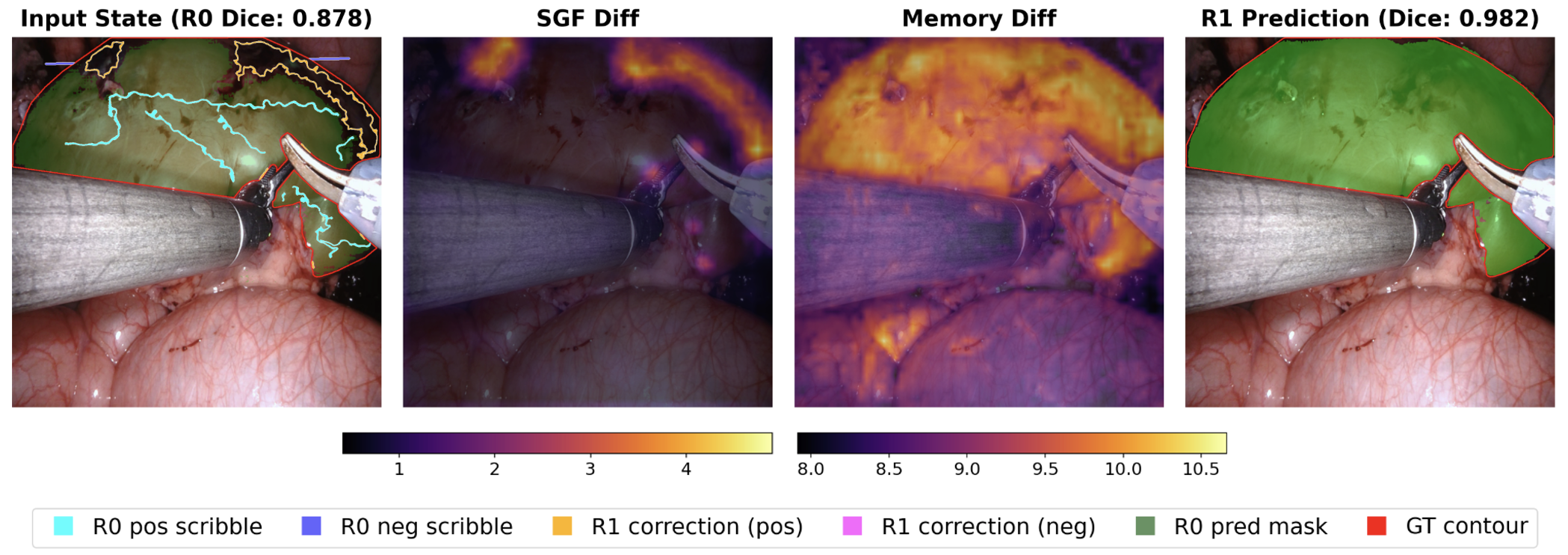}
    \caption{Qualitative visualization of feature changes from R0$\rightarrow$R1: SGF-induced query modification ($|F_q{-}F_{img}|$) and Memory-induced update ($|F_{mem}{-}F_{img}|$), alongside input scribbles and the refined R1 prediction.}
    \label{fig:component_ablation_demo}
\end{figure}

\noindent\textbf{Component contribution.}
Table~\ref{tab:component_ablation} indicates that SGF yields consistent improvements over the baseline by injecting the latest correction into the memory-attention query, while Memory further boosts performance and its benefit becomes more evident in later rounds as correction history accumulates.
Fig.~\ref{fig:component_ablation_demo} provides a qualitative view of these effects: SGF spatially diffuses the R1 correction scribble into the query features, while Memory Attention attends to the previous-round prediction stored in the memory bank and enhances features over relevant regions, leading to a refined mask.

\begin{table}[t]
\centering
\caption{Component ablation on \dataseta{}. All metrics (\%) are sample-average (m) and class-average (c). Each row adds one module to the previous configuration.}
\label{tab:component_ablation}
\resizebox{\textwidth}{!}{%
\setlength{\tabcolsep}{3.5pt}
\scriptsize
\begin{tabular}{l cccc cccc cccc}
\toprule
\multirow{2}{*}{Configuration}
& \multicolumn{4}{c}{\textbf{R0}}
& \multicolumn{4}{c}{\textbf{R1}}
& \multicolumn{4}{c}{\textbf{R2}} \\
\cmidrule(lr){2-5}\cmidrule(lr){6-9}\cmidrule(lr){10-13}
& mIoU & mDice & cIoU & cDice & mIoU & mDice & cIoU & cDice & mIoU & mDice & cIoU & cDice \\
\midrule
Baseline          & 69.88 & 80.73 & 69.68 & 80.51 & 76.56 & 85.65 & 75.56 & 84.86 & 80.21 & 88.16 & 78.56 & 86.94 \\
+ SGF             & 74.63 & 84.29 & 70.51 & 80.98 & 80.28 & 88.35 & 78.48 & 87.08 & 83.59 & 90.53 & 80.98 & 88.68 \\
+ SGF + Memory    & \textbf{75.77} & \textbf{85.06} & \textbf{71.97} & \textbf{82.09} & \textbf{85.96} & \textbf{92.03} & \textbf{82.43} & \textbf{89.47} & \textbf{88.30} & \textbf{93.49} & \textbf{84.90} & \textbf{91.18} \\
\bottomrule
\end{tabular}}
\vspace{-10pt}
\end{table}

\section{Conclusion}
\label{sec:conclusion}

We presented \ours{}, a scribble-promptable framework for interactive surgical segmentation. All added modules attach through standard embedding interfaces, enabling transfer to architectures such as \samthree{}. Experiments on \dataseta{} and the out-of-distribution \datasetb{} show that scribble prompts converge faster and more accurately than point or box. Future work will extend \ours{} to video-level annotation and validate usability with human annotators.


\clearpage
\appendix

\section{Iterative Refinement Algorithm}
\label{app:algorithm}

Algorithm~\ref{alg:iterative} provides the complete pseudocode for the iterative refinement procedure described in Sec.~\ref{sec:iterative}. At each round the dual-track scribble encoder processes both the accumulated and the latest correction scribble, while the memory bank retains the previous round's prediction to guide subsequent refinement.

\begin{algorithm}[H]
\caption{Iterative refinement with dual-track scribbles}
\label{alg:iterative}
\begin{algorithmic}[1]
\STATE \textbf{Input:} image $I$, initial scribble $S_0$, rounds $T$
\STATE $F_{\text{img}} \leftarrow \texttt{ImageEncoder}(I)$ \hfill (cached; frozen)
\STATE $\mathcal{M} \leftarrow \emptyset$ \hfill (memory bank size 1)
\STATE $S_{\text{acc}} \leftarrow S_0$
\FOR{$t=0$ to $T-1$}
    \STATE $E_{\text{acc}} \leftarrow \texttt{ScribbleEnc}(S_{\text{acc}})$ \hfill (Track~1)
    \STATE $E_t \leftarrow \texttt{ScribbleEnc}(S_t)$ \hfill (Track~2; latest)
    \STATE $E_{\text{mask}} \leftarrow 
    \begin{cases}
        \texttt{MaskEncoder}(\hat{M}_{t-1}), & t>0 \\
        \emptyset, & t=0
    \end{cases}$ \hfill (prev. mask prior)
    \STATE $F_q \leftarrow \texttt{SGF}(F_{\text{img}}, E_t)$
    \STATE $F_{\text{mem}} \leftarrow \texttt{MemAttn}(F_q, \mathcal{M})$ \hfill (no-mem embedding if $\mathcal{M}=\emptyset$)
    \STATE $\hat{M}_t \leftarrow \texttt{MaskDecoder}(F_{\text{mem}}, E_{\text{acc}} + E_{\text{mask}})$
    \STATE $\mathcal{M} \leftarrow \texttt{MemEncode}(F_{\text{img}}, \hat{M}_t)$
    \STATE Obtain next correction scribble $S_{t+1}$ from user; update $S_{\text{acc}} \leftarrow \max(S_{\text{acc}}, S_{t+1})$
\ENDFOR
\STATE \textbf{Output:} refined mask $\hat{M}_{T-1}$
\end{algorithmic}
\end{algorithm}

\section{Additional Comparison Tables}
\label{app:additional_tables}

We provide extended comparison tables that complement the main results in Table~\ref{tab:main_allrounds}. Table~\ref{tab:classavg_r0r2r4} reports class-average metrics on \dataseta{}, while Tables~\ref{tab:ood_main_allrounds} and~\ref{tab:ood_classavg_r012} give sample-average and class-average breakdowns on \datasetb{} (OOD), respectively.

\begin{table}[H]
\centering
\caption{Class-average comparison on \dataseta{} (EndoVis 2018 test). All metrics (\%) are averaged over $C{=}10$ classes. $R{k}$ denotes the prediction after $k$ refinement rounds. $\Delta$ reports the cIoU gain from $R0$ to $R4$.}
\label{tab:classavg_r0r2r4}
\setlength{\tabcolsep}{3pt}
\scriptsize
\resizebox{\linewidth}{!}{%
\begin{tabular}{l cc cc cc c}
\toprule
\multirow{2}{*}{Method}
& \multicolumn{2}{c}{\textbf{R0}}
& \multicolumn{2}{c}{\textbf{R2}}
& \multicolumn{2}{c}{\textbf{R4}}
& \multicolumn{1}{c}{$\Delta$cIoU} \\
\cmidrule(lr){2-3}\cmidrule(lr){4-5}\cmidrule(lr){6-7}
& cIoU & cDice & cIoU & cDice & cIoU & cDice & \\
\midrule
SCISSR (Adaptive)          & 72.08 & 82.18 & 84.84 & 91.09 & 86.89 & 92.47 & $+$14.81 \\
SCISSR (Contour only)      & 74.66 & 83.93 & 86.31 & 92.06 & \textbf{87.92} & \textbf{93.05} & $+$13.26 \\
SCISSR (Wave only)         & 69.74 & 80.36 & 84.99 & 91.19 & 86.67 & 92.26 & $+$16.93 \\
SCISSR (Centerline only)   & 71.37 & 81.82 & 82.20 & 89.41 & 83.29 & 90.08 & $+$11.92 \\
\midrule
SAM2 Tiny (1pt/CC)               & 54.68 & 65.83 & 60.30 & 69.86 & 54.69 & 64.17 & $+$0.01 \\
SAM3 (1pt/CC)                    & 53.28 & 64.36 & 57.82 & 67.05 & 51.89 & 59.83 & $-$1.39 \\
\bottomrule
\end{tabular}%
}
\end{table}

\begin{table}[H]
\centering
\caption{Comparison on \datasetb{} (CholecSeg8k test, OOD). Sample-average metrics (\%) over $N{=}8{,}800$ test masks. $R{k}$ denotes the prediction after $k$ refinement rounds. $\Delta$ reports the mIoU gain from $R0$ to $R2$.}
\label{tab:ood_main_allrounds}
\setlength{\tabcolsep}{3pt}
\scriptsize
\resizebox{\linewidth}{!}{%
\begin{tabular}{l cc cc cc c}
\toprule
\multirow{2}{*}{Method}
& \multicolumn{2}{c}{\textbf{R0}}
& \multicolumn{2}{c}{\textbf{R1}}
& \multicolumn{2}{c}{\textbf{R2}}
& \multirow{2}{*}{$\Delta$mIoU} \\
\cmidrule(lr){2-3}\cmidrule(lr){4-5}\cmidrule(lr){6-7}
& mIoU & mDice & mIoU & mDice & mIoU & mDice & \\
\midrule
SCISSR (Adaptive)          & 83.42 & 90.24 & 90.87 & 94.98 & \textbf{92.30} & \textbf{95.82} & $+$8.88 \\
\midrule
SAM2 Tiny (1pt/CC)               & 62.18 & 72.30 & 63.09 & 74.33 & 62.18 & 73.44 & $+$0.00 \\
SAM3 (1pt/CC)                    & 56.74 & 67.24 & 58.10 & 68.25 & 52.62 & 61.69 & $-$4.12 \\
\midrule
\samtwo{} Tiny (BBox)            & 76.22 & 84.62 & \multicolumn{2}{c}{--} & \multicolumn{2}{c}{--} & -- \\
\samthree{} (BBox)               & 77.85 & 85.78 & \multicolumn{2}{c}{--} & \multicolumn{2}{c}{--} & -- \\
MedSAM2 (BBox)                   & 76.74 & 82.69 & \multicolumn{2}{c}{--} & \multicolumn{2}{c}{--} & -- \\
\bottomrule
\end{tabular}%
}
\end{table}

\begin{table}[H]
\centering
\caption{Comparison on \datasetb{} (CholecSeg8k test, OOD). Class-average metrics (\%) over $C{=}12$ classes. $\Delta$ reports the cIoU gain from $R0$ to $R2$.}
\label{tab:ood_classavg_r012}
\setlength{\tabcolsep}{3pt}
\scriptsize
\resizebox{\linewidth}{!}{%
\begin{tabular}{l cc cc cc c}
\toprule
\multirow{2}{*}{Method}
& \multicolumn{2}{c}{\textbf{R0}}
& \multicolumn{2}{c}{\textbf{R1}}
& \multicolumn{2}{c}{\textbf{R2}}
& \multirow{2}{*}{$\Delta$cIoU} \\
\cmidrule(lr){2-3}\cmidrule(lr){4-5}\cmidrule(lr){6-7}
& cIoU & cDice & cIoU & cDice & cIoU & cDice & \\
\midrule
SCISSR (Adaptive)        & 80.58 & 87.96 & 87.63 & 92.77 & \textbf{89.32} & \textbf{93.79} & $+$8.74 \\
\midrule
SAM2 Tiny (1pt/CC)             & 59.23 & 69.47 & 64.38 & 74.55 & 65.79 & 75.61 & $+$6.56 \\
SAM3 (1pt/CC)                  & 58.71 & 67.96 & 61.75 & 70.92 & 56.27 & 65.91 & $-$2.44 \\
\midrule
\samtwo{} Tiny (BBox)          & 76.01 & 84.08 & \multicolumn{2}{c}{--} & \multicolumn{2}{c}{--} & -- \\
\samthree{} (BBox)             & 75.68 & 83.76 & \multicolumn{2}{c}{--} & \multicolumn{2}{c}{--} & -- \\
MedSAM2 (BBox)                 & 75.36 & 81.35 & \multicolumn{2}{c}{--} & \multicolumn{2}{c}{--} & -- \\
\bottomrule
\end{tabular}%
}
\end{table}

\section{Per-Class Detailed Results}
\label{app:per_class}

Tables~\ref{tab:per_class_allrounds}--\ref{tab:per_class_centerline} report per-class IoU and Dice for each scribble strategy on \dataseta{} across all five rounds. Table~\ref{tab:per_class_endovis_baselines} lists the corresponding per-class baseline numbers. Table~\ref{tab:component_ablation_perclass} breaks down the component ablation by class. Tables~\ref{tab:per_class_cholecseg8k_sam2_10pts}--\ref{tab:per_class_cholecseg8k_baselines} present per-class results on \datasetb{} (OOD), and Table~\ref{tab:convergence_ood} reports convergence efficiency on \datasetb{}.

\begin{table}[H]
\centering
\caption{Per-class results of \ours{} on \dataseta{} (contour strategy). IoU and Dice (\%) from Round~0 to Round~4. $\Delta$ denotes the IoU gain from R0 to R4.}
\label{tab:per_class_allrounds}
\setlength{\tabcolsep}{3pt}
\scriptsize
\resizebox{\linewidth}{!}{%
\begin{tabular}{l r cc cc cc cc cc r}
\toprule
\multirow{2}{*}{Class} & \multirow{2}{*}{$N$}
& \multicolumn{2}{c}{\textbf{R0}}
& \multicolumn{2}{c}{\textbf{R1}}
& \multicolumn{2}{c}{\textbf{R2}}
& \multicolumn{2}{c}{\textbf{R3}}
& \multicolumn{2}{c}{\textbf{R4}}
& \multirow{2}{*}{$\Delta$IoU} \\
\cmidrule(lr){3-4}\cmidrule(lr){5-6}\cmidrule(lr){7-8}\cmidrule(lr){9-10}\cmidrule(lr){11-12}
 & & IoU & Dice & IoU & Dice & IoU & Dice & IoU & Dice & IoU & Dice & \\
\midrule
\multicolumn{13}{l}{\textit{Instruments}} \\
\quad Shaft              & 843 & 90.71 & 94.94 & 93.91 & 96.83 & 94.28 & 96.99 & 95.16 & 97.50 & 95.42 & 97.63 & $+$4.71 \\
\quad Clasper            & 875 & 74.62 & 85.06 & 82.30 & 90.17 & 83.89 & 91.14 & 85.27 & 91.95 & 86.06 & 92.42 & $+$11.44 \\
\quad Wrist              & 822 & 76.76 & 86.07 & 88.15 & 93.47 & 88.88 & 93.93 & 90.31 & 94.76 & 90.92 & 95.10 & $+$14.16 \\
\midrule
\multicolumn{13}{l}{\textit{Tissues}} \\
\quad Kidney parenchyma  & 949 & 82.40 & 89.42 & 93.25 & 96.44 & 94.27 & 97.00 & 94.95 & 97.37 & 95.39 & 97.61 & $+$12.99 \\
\quad Covered kidney     & 484 & 73.65 & 84.31 & 92.86 & 96.27 & 94.48 & 97.15 & 95.13 & 97.50 & 95.80 & 97.85 & $+$22.15 \\
\quad Small intestine    & 225 & 90.73 & 94.94 & 94.02 & 96.84 & 94.57 & 97.14 & 95.26 & 97.52 & 95.85 & 97.85 & $+$5.12 \\
\midrule
\multicolumn{13}{l}{\textit{Other}} \\
\quad Thread             & 102 & 57.05 & 72.22 & 67.78 & 80.59 & 68.01 & 80.74 & 68.25 & 80.89 & 68.29 & 80.89 & $+$11.24 \\
\quad Clamps             &  69 & 75.72 & 85.47 & 83.75 & 90.81 & 86.93 & 92.86 & 88.74 & 93.97 & 88.48 & 93.84 & $+$12.76 \\
\quad Suturing needle    &  95 & 44.54 & 58.41 & 65.45 & 77.29 & 65.99 & 77.98 & 67.95 & 79.44 & 69.46 & 80.72 & $+$24.92 \\
\quad Ultrasound probe   & 152 & 80.41 & 88.50 & 89.97 & 94.61 & 91.79 & 95.65 & 93.24 & 96.45 & 93.54 & 96.58 & $+$13.13 \\
\bottomrule
\end{tabular}%
}
\end{table}

\begin{table}[H]
\centering
\caption{Per-class results of \ours{} on \dataseta{} (adaptive strategy). IoU and Dice (\%) from Round~0 to Round~4. $\Delta$ denotes the IoU gain from R0 to R4.}
\label{tab:per_class_adaptive}
\setlength{\tabcolsep}{3pt}
\scriptsize
\resizebox{\linewidth}{!}{%
\begin{tabular}{l r cc cc cc cc cc r}
\toprule
\multirow{2}{*}{Class} & \multirow{2}{*}{$N$}
& \multicolumn{2}{c}{\textbf{R0}}
& \multicolumn{2}{c}{\textbf{R1}}
& \multicolumn{2}{c}{\textbf{R2}}
& \multicolumn{2}{c}{\textbf{R3}}
& \multicolumn{2}{c}{\textbf{R4}}
& \multirow{2}{*}{$\Delta$IoU} \\
\cmidrule(lr){3-4}\cmidrule(lr){5-6}\cmidrule(lr){7-8}\cmidrule(lr){9-10}\cmidrule(lr){11-12}
 & & IoU & Dice & IoU & Dice & IoU & Dice & IoU & Dice & IoU & Dice & \\
\midrule
\multicolumn{13}{l}{\textit{Instruments}} \\
\quad Shaft              & 843 & 86.38 & 92.39 & 92.13 & 95.83 & 93.58 & 96.64 & 94.20 & 96.98 & 94.50 & 97.14 & $+$8.12 \\
\quad Clasper            & 875 & 71.43 & 82.88 & 81.37 & 89.58 & 83.43 & 90.84 & 84.41 & 91.41 & 85.26 & 91.95 & $+$13.83 \\
\quad Wrist              & 822 & 70.12 & 81.35 & 84.43 & 91.14 & 87.46 & 93.03 & 88.68 & 93.75 & 89.57 & 94.31 & $+$19.45 \\
\midrule
\multicolumn{13}{l}{\textit{Tissues}} \\
\quad Kidney parenchyma  & 949 & 79.86 & 87.83 & 90.01 & 94.56 & 91.88 & 95.62 & 92.76 & 96.12 & 93.32 & 96.43 & $+$13.46 \\
\quad Covered kidney     & 484 & 70.28 & 81.91 & 83.18 & 90.38 & 88.37 & 93.57 & 90.59 & 94.86 & 92.25 & 95.81 & $+$21.97 \\
\quad Small intestine    & 225 & 85.73 & 91.99 & 92.08 & 95.76 & 93.31 & 96.45 & 94.28 & 97.00 & 94.92 & 97.33 & $+$9.19 \\
\midrule
\multicolumn{13}{l}{\textit{Other}} \\
\quad Thread             & 102 & 58.52 & 73.45 & 67.79 & 80.62 & 68.62 & 81.20 & 68.83 & 81.35 & 68.80 & 81.32 & $+$10.28 \\
\quad Clamps             &  69 & 73.22 & 82.66 & 82.85 & 89.34 & 84.16 & 90.12 & 85.27 & 91.10 & 87.40 & 92.92 & $+$14.18 \\
\quad Suturing needle    &  95 & 45.31 & 59.14 & 62.76 & 74.82 & 66.59 & 78.26 & 69.04 & 80.51 & 70.15 & 81.35 & $+$24.84 \\
\quad Ultrasound probe   & 152 & 79.95 & 88.21 & 89.00 & 93.98 & 91.04 & 95.19 & 92.18 & 95.85 & 92.76 & 96.18 & $+$12.81 \\
\bottomrule
\end{tabular}%
}
\end{table}

\begin{table}[H]
\centering
\caption{Per-class results of \ours{} on \dataseta{} (wave-only strategy). IoU and Dice (\%) from Round~0 to Round~4. $\Delta$ denotes the IoU gain from R0 to R4.}
\label{tab:per_class_wave}
\setlength{\tabcolsep}{3pt}
\scriptsize
\resizebox{\linewidth}{!}{%
\begin{tabular}{l r cc cc cc cc cc r}
\toprule
\multirow{2}{*}{Class} & \multirow{2}{*}{$N$}
& \multicolumn{2}{c}{\textbf{R0}}
& \multicolumn{2}{c}{\textbf{R1}}
& \multicolumn{2}{c}{\textbf{R2}}
& \multicolumn{2}{c}{\textbf{R3}}
& \multicolumn{2}{c}{\textbf{R4}}
& \multirow{2}{*}{$\Delta$IoU} \\
\cmidrule(lr){3-4}\cmidrule(lr){5-6}\cmidrule(lr){7-8}\cmidrule(lr){9-10}\cmidrule(lr){11-12}
 & & IoU & Dice & IoU & Dice & IoU & Dice & IoU & Dice & IoU & Dice & \\
\midrule
\multicolumn{13}{l}{\textit{Instruments}} \\
\quad Shaft              & 843 & 84.54 & 91.30 & 91.91 & 95.72 & 93.45 & 96.57 & 94.04 & 96.88 & 94.46 & 97.10 & $+$9.92 \\
\quad Clasper            & 875 & 68.72 & 80.98 & 80.09 & 88.74 & 82.89 & 90.53 & 84.19 & 91.33 & 84.77 & 91.66 & $+$16.05 \\
\quad Wrist              & 822 & 67.49 & 79.49 & 83.63 & 90.60 & 87.26 & 92.91 & 88.36 & 93.58 & 89.01 & 93.98 & $+$21.52 \\
\midrule
\multicolumn{13}{l}{\textit{Tissues}} \\
\quad Kidney parenchyma  & 949 & 78.54 & 86.99 & 90.23 & 94.74 & 92.82 & 96.21 & 93.74 & 96.72 & 94.26 & 97.00 & $+$15.72 \\
\quad Covered kidney     & 484 & 67.40 & 79.78 & 88.14 & 93.64 & 91.49 & 95.52 & 93.58 & 96.66 & 94.58 & 97.19 & $+$27.18 \\
\quad Small intestine    & 225 & 84.89 & 91.55 & 92.78 & 96.17 & 94.18 & 96.94 & 94.74 & 97.25 & 95.15 & 97.48 & $+$10.26 \\
\midrule
\multicolumn{13}{l}{\textit{Other}} \\
\quad Thread             & 102 & 50.83 & 66.80 & 63.14 & 77.07 & 66.98 & 80.04 & 68.26 & 80.97 & 68.81 & 81.35 & $+$17.98 \\
\quad Clamps             &  69 & 70.62 & 79.96 & 81.22 & 88.24 & 83.24 & 90.05 & 83.87 & 90.47 & 84.12 & 90.62 & $+$13.50 \\
\quad Suturing needle    &  95 & 45.51 & 59.48 & 63.46 & 75.66 & 65.86 & 77.57 & 66.91 & 78.52 & 68.36 & 79.75 & $+$22.85 \\
\quad Ultrasound probe   & 152 & 78.82 & 87.24 & 89.84 & 94.51 & 91.77 & 95.60 & 92.73 & 96.19 & 93.21 & 96.44 & $+$14.39 \\
\bottomrule
\end{tabular}%
}
\end{table}

\begin{table}[H]
\centering
\caption{Per-class results of \ours{} on \dataseta{} (centerline-only strategy). IoU and Dice (\%) from Round~0 to Round~4. $\Delta$ denotes the IoU gain from R0 to R4.}
\label{tab:per_class_centerline}
\setlength{\tabcolsep}{3pt}
\scriptsize
\resizebox{\linewidth}{!}{%
\begin{tabular}{l r cc cc cc cc cc r}
\toprule
\multirow{2}{*}{Class} & \multirow{2}{*}{$N$}
& \multicolumn{2}{c}{\textbf{R0}}
& \multicolumn{2}{c}{\textbf{R1}}
& \multicolumn{2}{c}{\textbf{R2}}
& \multicolumn{2}{c}{\textbf{R3}}
& \multicolumn{2}{c}{\textbf{R4}}
& \multirow{2}{*}{$\Delta$IoU} \\
\cmidrule(lr){3-4}\cmidrule(lr){5-6}\cmidrule(lr){7-8}\cmidrule(lr){9-10}\cmidrule(lr){11-12}
 & & IoU & Dice & IoU & Dice & IoU & Dice & IoU & Dice & IoU & Dice & \\
\midrule
\multicolumn{13}{l}{\textit{Instruments}} \\
\quad Shaft              & 843 & 85.33 & 91.69 & 91.07 & 95.22 & 92.54 & 96.05 & 93.08 & 96.34 & 93.46 & 96.55 & $+$8.13 \\
\quad Clasper            & 875 & 71.03 & 82.58 & 79.54 & 88.30 & 80.85 & 89.08 & 81.30 & 89.33 & 81.45 & 89.41 & $+$10.42 \\
\quad Wrist              & 822 & 67.02 & 79.38 & 79.93 & 88.25 & 82.70 & 89.96 & 83.56 & 90.47 & 83.95 & 90.70 & $+$16.93 \\
\midrule
\multicolumn{13}{l}{\textit{Tissues}} \\
\quad Kidney parenchyma  & 949 & 77.85 & 86.46 & 88.23 & 93.52 & 89.75 & 94.35 & 90.36 & 94.67 & 90.65 & 94.81 & $+$12.80 \\
\quad Covered kidney     & 484 & 66.93 & 79.28 & 75.88 & 85.58 & 80.37 & 88.52 & 82.95 & 90.16 & 84.67 & 91.22 & $+$17.74 \\
\quad Small intestine    & 225 & 85.16 & 91.67 & 91.11 & 95.24 & 92.65 & 96.11 & 93.39 & 96.51 & 93.79 & 96.73 & $+$8.63 \\
\midrule
\multicolumn{13}{l}{\textit{Other}} \\
\quad Thread             & 102 & 59.01 & 73.83 & 67.74 & 80.54 & 68.59 & 81.19 & 68.95 & 81.44 & 68.79 & 81.33 & $+$9.78 \\
\quad Clamps             &  69 & 75.84 & 85.55 & 82.71 & 90.17 & 83.19 & 90.32 & 83.38 & 90.45 & 83.33 & 90.43 & $+$7.49 \\
\quad Suturing needle    &  95 & 45.86 & 59.74 & 60.05 & 72.52 & 62.94 & 74.99 & 63.68 & 75.62 & 64.06 & 76.02 & $+$18.20 \\
\quad Ultrasound probe   & 152 & 79.68 & 88.05 & 87.53 & 92.97 & 88.45 & 93.47 & 88.65 & 93.58 & 88.76 & 93.64 & $+$9.08 \\
\bottomrule
\end{tabular}%
}
\end{table}

\begin{table}[H]
\centering
\caption{Per-class results of baselines on \dataseta{}. IoU (\%) from Round~0 to Round~4 for iterative methods; single-round IoU and Dice for box methods.}
\label{tab:per_class_endovis_baselines}
\setlength{\tabcolsep}{2.5pt}
\scriptsize
\resizebox{\linewidth}{!}{%
\begin{tabular}{l r ccccc ccccc cc cc cc}
\toprule
\multirow{2}{*}{Class} & \multirow{2}{*}{$N$}
& \multicolumn{5}{c}{SAM2 Tiny (1pt/CC) IoU}
& \multicolumn{5}{c}{SAM3 (1pt/CC) IoU}
& \multicolumn{2}{c}{\samtwo{} Tiny (BBox)}
& \multicolumn{2}{c}{\samthree{} (BBox)}
& \multicolumn{2}{c}{MedSAM2 (BBox)} \\
\cmidrule(lr){3-7}\cmidrule(lr){8-12}\cmidrule(lr){13-14}\cmidrule(lr){15-16}\cmidrule(lr){17-18}
& & \textbf{R0} & \textbf{R1} & \textbf{R2} & \textbf{R3} & \textbf{R4} & \textbf{R0} & \textbf{R1} & \textbf{R2} & \textbf{R3} & \textbf{R4} & IoU & Dice & IoU & Dice & IoU & Dice \\
\midrule
\multicolumn{18}{l}{\textit{Instruments}} \\
\quad Shaft              & 843 & 70.52 & 77.03 & 77.06 & 73.32 & 67.08 & 69.99 & 74.09 & 70.40 & 62.25 & 53.81 & 64.98 & 73.19 & 67.11 & 74.53 & 72.78 & 82.25 \\
\quad Clasper            & 875 & 44.65 & 54.05 & 52.87 & 46.95 & 39.08 & 55.54 & 57.10 & 54.82 & 49.19 & 42.66 & 45.04 & 55.58 & 58.17 & 68.72 & 52.58 & 67.12 \\
\quad Wrist              & 822 & 54.56 & 65.58 & 70.06 & 71.21 & 70.36 & 55.57 & 63.38 & 69.02 & 71.80 & 72.10 & 67.54 & 76.50 & 71.29 & 80.07 & 65.30 & 76.22 \\
\midrule
\multicolumn{18}{l}{\textit{Tissues}} \\
\quad Kidney parenchyma  & 949 & 63.64 & 63.38 & 60.87 & 58.56 & 56.21 & 61.22 & 58.96 & 53.66 & 48.45 & 43.72 & 76.59 & 85.00 & 80.58 & 88.18 & 15.02 & 19.40 \\
\quad Covered kidney     & 484 & 43.86 & 47.46 & 45.83 & 41.97 & 38.53 & 39.95 & 44.23 & 42.69 & 39.07 & 35.37 & 59.78 & 71.95 & 59.86 & 72.35 & 10.11 & 13.57 \\
\quad Small intestine    & 225 & 65.73 & 81.15 & 82.96 & 80.47 & 76.34 & 64.35 & 79.16 & 80.16 & 73.53 & 66.99 & 87.40 & 92.45 & 88.67 & 93.60 & 50.04 & 57.10 \\
\midrule
\multicolumn{18}{l}{\textit{Other}} \\
\quad Thread             & 102 & 22.48 &  9.77 &  5.82 &  3.44 &  2.29 &  8.87 &  6.15 &  2.54 &  1.59 &  1.47 & 17.72 & 26.18 & 32.46 & 44.27 &  1.43 &  2.56 \\
\quad Clamps             &  69 & 83.38 & 87.07 & 87.49 & 87.58 & 87.18 & 77.51 & 82.54 & 85.44 & 86.26 & 86.65 & 86.86 & 92.60 & 87.21 & 92.82 & 75.74 & 85.56 \\
\quad Suturing needle    &  95 & 35.79 & 42.88 & 47.78 & 51.68 & 51.96 & 38.10 & 41.75 & 41.80 & 44.56 & 43.76 & 50.63 & 63.35 & 57.79 & 69.95 & 27.67 & 38.60 \\
\quad Ultrasound probe   & 152 & 62.19 & 71.77 & 72.22 & 67.52 & 57.88 & 61.68 & 74.85 & 77.64 & 77.31 & 72.34 & 81.23 & 88.53 & 84.33 & 90.69 & 80.80 & 88.03 \\
\bottomrule
\end{tabular}%
}
\end{table}

\begin{table}[H]
\centering
\caption{Component ablation per-class on \dataseta{}. IoU (\%) at R0 and R2 for each configuration. $\Delta$ denotes the IoU gain from Baseline to SGF+Memory at R2.}
\label{tab:component_ablation_perclass}
\setlength{\tabcolsep}{3pt}
\scriptsize
\resizebox{\linewidth}{!}{%
\begin{tabular}{l r cc cc cc r}
\toprule
\multirow{2}{*}{Class} & \multirow{2}{*}{$N$}
& \multicolumn{2}{c}{Baseline}
& \multicolumn{2}{c}{+ SGF}
& \multicolumn{2}{c}{+ SGF + Memory}
& \multirow{2}{*}{$\Delta$IoU} \\
\cmidrule(lr){3-4}\cmidrule(lr){5-6}\cmidrule(lr){7-8}
& & \textbf{R0} & \textbf{R2} & \textbf{R0} & \textbf{R2} & \textbf{R0} & \textbf{R2} & \\
\midrule
\multicolumn{9}{l}{\textit{Instruments}} \\
\quad Shaft              & 843 & 81.75 & 89.02 & 87.20 & 91.53 & 86.22 & 93.40 & $+$4.38 \\
\quad Clasper            & 875 & 60.01 & 73.76 & 70.19 & 78.16 & 71.11 & 83.39 & $+$9.63 \\
\quad Wrist              & 822 & 59.64 & 73.35 & 66.26 & 78.55 & 69.96 & 87.55 & $+$14.20 \\
\midrule
\multicolumn{9}{l}{\textit{Tissues}} \\
\quad Kidney parenchyma  & 949 & 77.26 & 84.75 & 79.92 & 88.12 & 79.83 & 91.20 & $+$6.45 \\
\quad Covered kidney     & 484 & 65.36 & 78.98 & 69.63 & 81.90 & 69.78 & 87.92 & $+$8.94 \\
\quad Small intestine    & 225 & 83.45 & 90.78 & 85.44 & 92.24 & 87.04 & 93.72 & $+$2.94 \\
\midrule
\multicolumn{9}{l}{\textit{Other}} \\
\quad Thread             & 102 & 63.11 & 66.52 & 55.63 & 65.69 & 57.56 & 68.45 & $+$1.93 \\
\quad Clamps             &  69 & 76.53 & 82.02 & 73.35 & 83.76 & 74.22 & 85.29 & $+$3.27 \\
\quad Suturing needle    &  95 & 53.19 & 60.12 & 40.08 & 62.13 & 44.65 & 66.19 & $+$6.07 \\
\quad Ultrasound probe   & 152 & 76.48 & 86.32 & 77.45 & 87.75 & 79.30 & 91.07 & $+$4.75 \\
\bottomrule
\end{tabular}%
}
\end{table}

\begin{table}[H]
\centering
\caption{Per-class results of SAM2 Tiny (10pt/ch) on \datasetb{} (OOD). IoU and Dice (\%) from Round~0 to Round~2. $\Delta$ denotes the IoU gain from R0 to R2.}
\label{tab:per_class_cholecseg8k_sam2_10pts}
\setlength{\tabcolsep}{3pt}
\scriptsize
\resizebox{\linewidth}{!}{%
\begin{tabular}{l r cc cc cc r}
\toprule
\multirow{2}{*}{Class} & \multirow{2}{*}{$N$}
& \multicolumn{2}{c}{\textbf{R0}}
& \multicolumn{2}{c}{\textbf{R1}}
& \multicolumn{2}{c}{\textbf{R2}}
& \multirow{2}{*}{$\Delta$IoU} \\
\cmidrule(lr){3-4}\cmidrule(lr){5-6}\cmidrule(lr){7-8}
& & IoU & Dice & IoU & Dice & IoU & Dice & \\
\midrule
\multicolumn{9}{l}{\textit{Tissues / Organs}} \\
\quad Abdominal Wall          & 1464 & 71.44 & 81.60 & 80.96 & 88.12 & 81.92 & 88.52 & $+$10.48 \\
\quad Liver                   & 1679 & 67.43 & 78.46 & 72.06 & 82.18 & 74.49 & 83.83 & $+$7.06 \\
\quad Gastrointestinal Tract  &  941 & 61.84 & 71.81 & 68.16 & 77.26 & 70.54 & 78.92 & $+$8.70 \\
\quad Fat                     & 1519 & 56.59 & 68.70 & 70.36 & 80.07 & 73.18 & 81.45 & $+$16.59 \\
\quad Connective Tissue       &  240 & 53.89 & 67.37 & 57.54 & 71.23 & 62.34 & 75.21 & $+$8.45 \\
\quad Gallbladder             & 1279 & 70.91 & 80.03 & 72.94 & 81.64 & 75.71 & 83.71 & $+$4.80 \\
\quad Hepatic Vein            &   77 & 45.24 & 60.46 & 39.12 & 52.83 & 28.52 & 40.06 & $-$16.72 \\
\quad Liver Ligament          &   80 & 94.63 & 97.21 & 95.46 & 97.67 & 95.36 & 97.62 & $+$0.73 \\
\quad Cystic Duct             &    1 & 96.75 & 98.35 & 95.02 & 97.45 & 95.90 & 97.91 & $-$0.85 \\
\midrule
\multicolumn{9}{l}{\textit{Instruments}} \\
\quad Grasper                 & 1080 & 78.52 & 87.35 & 76.46 & 85.52 & 75.80 & 84.19 & $-$2.72 \\
\quad L-hook Electrocautery   &  360 & 89.22 & 93.87 & 88.57 & 93.29 & 89.37 & 93.65 & $+$0.15 \\
\midrule
\multicolumn{9}{l}{\textit{Other}} \\
\quad Blood                   &   80 & 20.79 & 33.28 & 23.49 & 37.48 & 23.10 & 36.63 & $+$2.31 \\
\bottomrule
\end{tabular}%
}
\end{table}

\begin{table}[H]
\centering
\caption{Per-class results on \datasetb{} (Adaptive strategy, OOD). IoU and Dice (\%) from Round~0 to Round~2. $\Delta$ denotes the IoU gain from R0 to R2.}
\label{tab:per_class_cholecseg8k_r0r2}
\setlength{\tabcolsep}{3pt}
\scriptsize
\resizebox{\linewidth}{!}{%
\begin{tabular}{l r cc cc cc r}
\toprule
\multirow{2}{*}{Class} & \multirow{2}{*}{$N$}
& \multicolumn{2}{c}{\textbf{R0}}
& \multicolumn{2}{c}{\textbf{R1}}
& \multicolumn{2}{c}{\textbf{R2}}
& \multirow{2}{*}{$\Delta$IoU} \\
\cmidrule(lr){3-4}\cmidrule(lr){5-6}\cmidrule(lr){7-8}
& & IoU & Dice & IoU & Dice & IoU & Dice & \\
\midrule

\multicolumn{9}{l}{\textit{Tissues / Organs}} \\
\quad Abdominal Wall          & 1464 & 86.61 & 92.25 & 94.35 & 96.97 & 95.71 & 97.75 & $+$9.10 \\
\quad Liver                   & 1679 & 86.79 & 92.72 & 93.66 & 96.70 & 95.14 & 97.50 & $+$8.35 \\
\quad Gastrointestinal Tract  &  941 & 77.03 & 86.07 & 88.31 & 93.60 & 89.47 & 94.29 & $+$12.44 \\
\quad Fat                     & 1519 & 86.89 & 92.80 & 93.27 & 96.47 & 94.91 & 97.36 & $+$8.02 \\
\quad Connective Tissue       &  240 & 79.13 & 88.03 & 87.36 & 93.15 & 90.52 & 94.97 & $+$11.39 \\
\quad Gallbladder             & 1279 & 78.79 & 86.50 & 89.01 & 93.75 & 90.32 & 94.63 & $+$11.53 \\
\quad Hepatic Vein            &   77 & 41.00 & 57.49 & 54.34 & 69.50 & 58.58 & 72.33 & $+$17.58 \\
\quad Liver Ligament          &   80 & 96.61 & 98.28 & 97.38 & 98.67 & 97.74 & 98.85 & $+$1.13 \\
\quad Cystic Duct             &    1 & 96.26 & 98.10 & 97.82 & 98.90 & 98.12 & 99.05 & $+$1.86 \\

\midrule
\multicolumn{9}{l}{\textit{Instruments}} \\
\quad Grasper                 & 1080 & 82.26 & 89.97 & 86.07 & 92.35 & 87.33 & 93.11 & $+$5.07 \\
\quad L-hook Electrocautery   &  360 & 89.55 & 93.81 & 92.98 & 96.17 & 93.35 & 96.38 & $+$3.80 \\

\midrule
\multicolumn{9}{l}{\textit{Other}} \\
\quad Blood                   &   80 & 66.07 & 79.54 & 77.07 & 86.98 & 80.60 & 89.19 & $+$14.53 \\
\bottomrule
\end{tabular}%
}
\end{table}

\begin{table}[H]
\centering
\caption{Per-class results of baselines on \datasetb{} (OOD). IoU (\%) from Round~0 to Round~2 for iterative methods; single-round IoU for box methods. $\Delta$ denotes the IoU change from R0 to R2.}
\label{tab:per_class_cholecseg8k_baselines}
\setlength{\tabcolsep}{3pt}
\scriptsize
\resizebox{\linewidth}{!}{%
\begin{tabular}{l r ccc ccc cc cc cc}
\toprule
\multirow{2}{*}{Class} & \multirow{2}{*}{$N$}
& \multicolumn{3}{c}{SAM2 Tiny (1pt/CC)}
& \multicolumn{3}{c}{SAM3 (1pt/CC)}
& \multicolumn{2}{c}{\samtwo{} Tiny (BBox)}
& \multicolumn{2}{c}{\samthree{} (BBox)}
& \multicolumn{2}{c}{MedSAM2 (BBox)} \\
\cmidrule(lr){3-5}\cmidrule(lr){6-8}\cmidrule(lr){9-10}\cmidrule(lr){11-12}\cmidrule(lr){13-14}
& & \textbf{R0} & \textbf{R1} & \textbf{R2} & \textbf{R0} & \textbf{R1} & \textbf{R2} & IoU & Dice & IoU & Dice & IoU & Dice \\
\midrule
\multicolumn{14}{l}{\textit{Tissues / Organs}} \\
\quad Abdominal Wall          & 1464 & 66.15 & 63.21 & 61.81 & 51.20 & 46.91 & 35.26 & 81.38 & 88.35 & 85.08 & 90.85 & 80.39 & 85.81 \\
\quad Liver                   & 1679 & 62.88 & 54.35 & 47.93 & 51.05 & 46.67 & 32.51 & 72.76 & 82.44 & 75.80 & 84.83 & 73.61 & 80.41 \\
\quad Gastrointestinal Tract  &  941 & 56.67 & 63.12 & 69.92 & 61.20 & 69.80 & 72.35 & 82.04 & 89.53 & 82.58 & 90.09 & 85.82 & 91.81 \\
\quad Fat                     & 1519 & 49.51 & 54.14 & 52.09 & 42.76 & 43.92 & 38.56 & 64.36 & 75.15 & 65.65 & 76.27 & 60.48 & 65.83 \\
\quad Connective Tissue       &  240 & 47.45 & 53.42 & 51.39 & 45.16 & 54.38 & 54.43 & 72.24 & 83.07 & 56.68 & 71.28 & 88.21 & 93.59 \\
\quad Gallbladder             & 1279 & 67.78 & 70.00 & 69.24 & 66.06 & 71.45 & 70.98 & 82.34 & 88.81 & 82.38 & 88.24 & 87.80 & 93.05 \\
\quad Hepatic Vein            &   77 & 31.81 & 50.26 & 66.70 & 27.93 & 42.74 & 51.81 & 63.90 & 77.60 & 62.63 & 76.51 & 57.65 & 72.60 \\
\quad Liver Ligament          &   80 & 95.36 & 92.57 & 90.68 & 86.44 & 84.92 & 72.79 & 97.16 & 98.56 & 97.09 & 98.52 & 95.88 & 97.89 \\
\quad Cystic Duct             &    1 & 59.22 & 86.67 & 92.97 & 94.74 & 96.38 & 65.16 & 97.53 & 98.75 & 97.57 & 98.77 & 96.99 & 98.47 \\
\midrule
\multicolumn{14}{l}{\textit{Instruments}} \\
\quad Grasper                 & 1080 & 70.08 & 76.20 & 77.68 & 73.31 & 77.79 & 77.44 & 77.46 & 86.14 & 82.11 & 89.74 & 75.82 & 83.15 \\
\quad L-hook Electrocautery   &  360 & 85.79 & 89.43 & 90.30 & 85.90 & 88.21 & 87.60 & 91.77 & 95.61 & 92.24 & 95.88 & 91.68 & 95.58 \\
\midrule
\multicolumn{14}{l}{\textit{Other}} \\
\quad Blood                   &   80 & 18.01 & 19.22 & 18.80 & 18.72 & 17.88 & 16.37 & 29.15 & 45.00 & 28.36 & 44.09 &  9.97 & 18.01 \\
\bottomrule
\end{tabular}%
}
\end{table}

\begin{table}[H]
\centering
\caption{Convergence efficiency on \datasetb{} (CholecSeg8k, OOD). Success rate (\%) is the fraction of $N{=}8{,}800$ test masks reaching the Dice threshold within 4 rounds. Mean rounds is averaged over successful masks only.}
\label{tab:convergence_ood}
\setlength{\tabcolsep}{4pt}
\scriptsize
\begin{tabular}{l l cc ccccc}
\toprule
\multirow{2}{*}{Method} & \multirow{2}{*}{Dice $\geq$} & \multirow{2}{*}{Success} & \multirow{2}{*}{Mean Rnd} & \multicolumn{5}{c}{Cumulative \% by round} \\
\cmidrule(lr){5-9}
& & (\%) & & \textbf{R0} & \textbf{R1} & \textbf{R2} & \textbf{R3} & \textbf{R4} \\
\midrule
\multirow{3}{*}{\ours{} (Adaptive)}
 & 0.75 & \textbf{99.4} & \textbf{1.28} & 78.2 & 94.8 & 98.0 & 99.1 & 99.4 \\
 & 0.85 & \textbf{98.0} & \textbf{1.50} & 62.4 & 88.6 & 94.8 & 97.1 & 98.0 \\
 & 0.90 & \textbf{94.0} & \textbf{1.73} & 47.4 & 79.3 & 88.6 & 92.2 & 94.0 \\
\midrule
\multirow{3}{*}{SAM2 Tiny (1pt/CC)}
 & 0.75 & 79.9 & 1.37 & 61.8 & 72.5 & 77.0 & 79.1 & 79.9 \\
 & 0.85 & 64.8 & 1.51 & 44.8 & 56.2 & 61.1 & 63.7 & 64.8 \\
 & 0.90 & 49.8 & 1.61 & 32.9 & 41.3 & 46.2 & 48.6 & 49.8 \\
\midrule
\multirow{3}{*}{SAM3 (1pt/CC)}
 & 0.75 & 65.8 & 1.31 & 52.1 & 61.4 & 64.0 & 65.2 & 65.8 \\
 & 0.85 & 53.8 & 1.47 & 38.1 & 47.6 & 51.4 & 53.0 & 53.8 \\
 & 0.90 & 42.8 & 1.53 & 29.4 & 36.9 & 40.3 & 41.9 & 42.8 \\
\bottomrule
\end{tabular}
\end{table}

\section{Point Prompt Density Analysis}
\label{app:point_density}

To investigate whether simply increasing the number of point prompts can close the gap with scribble prompts, we vary the point density from 1 to 50 per channel on a subset of \datasetb{} (4 videos). As shown in Table~\ref{tab:point_degradation}, both \samtwo{} Tiny and \samthree{} peak at 10 points and degrade sharply beyond that, confirming that the point interface cannot exploit denser spatial signals.

\begin{table}[H]
\centering
\caption{Effect of point prompt density on \samtwo{} Tiny and \samthree{} (\datasetb{}, 4 videos, Round~2). Performance degrades sharply beyond 10 points per channel.}
\label{tab:point_degradation}
\setlength{\tabcolsep}{4.5pt}
\begin{tabular}{l cc cc cc cc}
\toprule
\multirow{2}{*}{Method} & \multicolumn{2}{c}{1\,pt} & \multicolumn{2}{c}{10\,pts} & \multicolumn{2}{c}{30\,pts} & \multicolumn{2}{c}{50\,pts} \\
\cmidrule(lr){2-3} \cmidrule(lr){4-5} \cmidrule(lr){6-7} \cmidrule(lr){8-9}
 & IoU & Dice & IoU & Dice & IoU & Dice & IoU & Dice \\
\midrule
\samtwo{} Tiny & 60.25 & 71.79 & \textbf{73.76} & \textbf{82.49} & 34.36 & 44.24 & 22.43 & 30.84 \\
\samthree{}     & 50.29 & 60.14 & 61.46 & 72.79 &  1.12 &  1.55 &  0.00 &  0.00 \\
\bottomrule
\end{tabular}
\end{table}

\section{Scribble Strategy Ablation}
\label{app:strategy_comparison}

\begin{figure}[H]
    \centering
    \includegraphics[width=0.7\linewidth]{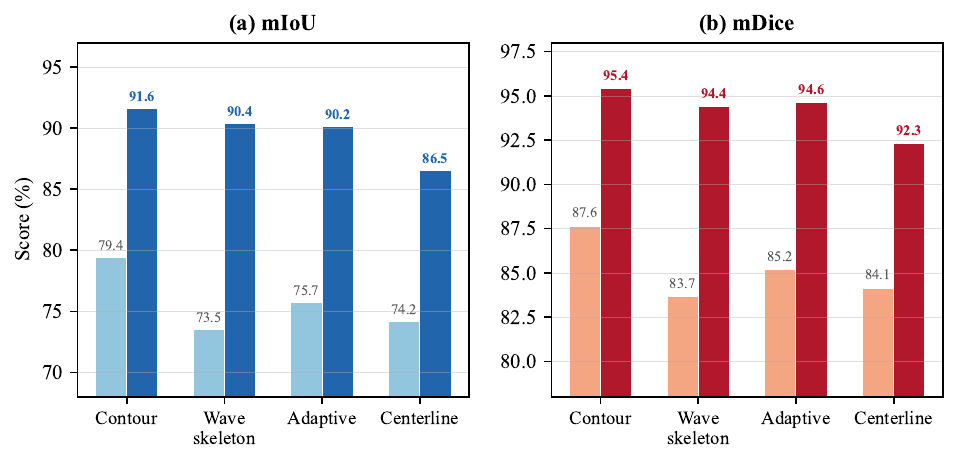}\\[2pt]
    \includegraphics[width=0.45\linewidth]{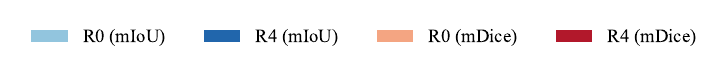}
    \caption{Scribble strategy ablation on \dataseta{}. Contour achieves the best R0 and R4 across both mIoU and mDice.}
    \label{fig:strategy_comparison}
\end{figure}

Fig.~\ref{fig:strategy_comparison} visualizes the four scribble generation strategies evaluated in Sec.~\ref{sec:ablation}. Contour-based scribbles consistently achieve the highest mIoU and mDice at both R0 and R4, while centerline-only lags due to the absence of boundary cues.

\section{Additional Figures}
\label{app:additional_figures}

\begin{figure}[H]
    \centering
    \includegraphics[width=0.85\linewidth]{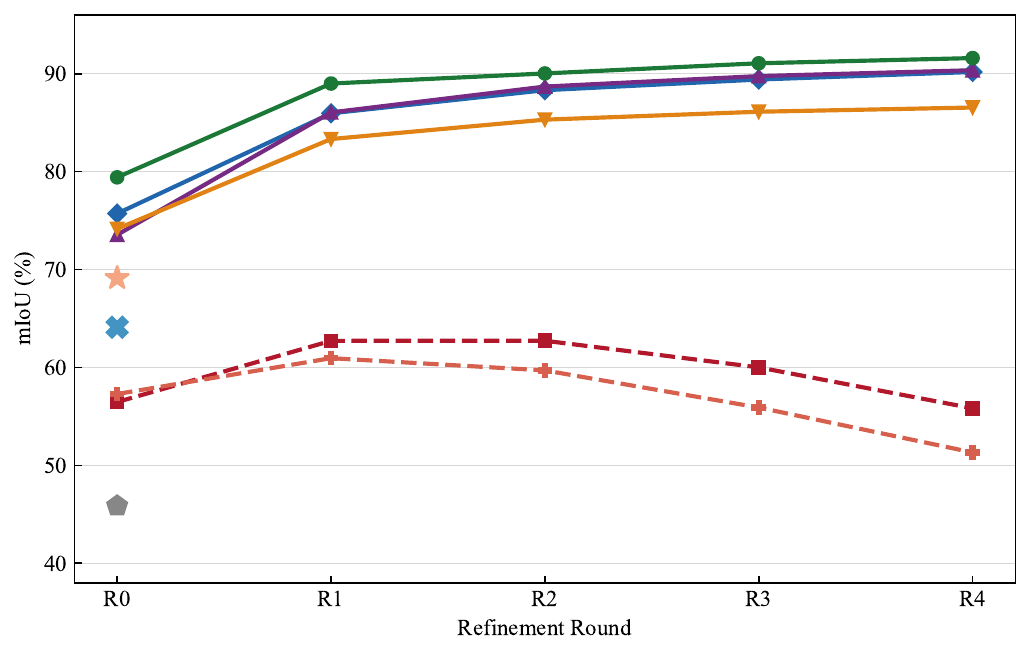}\\[2pt]
    \includegraphics[width=0.6\linewidth]{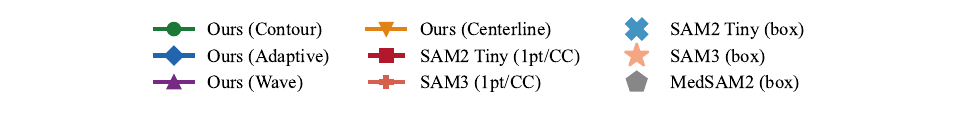}
    \caption{Iterative refinement curves on \dataseta{}. This figure visualizes the same mIoU data reported in Table~\ref{tab:main_allrounds}. All four scribble strategies improve steadily across rounds, while point-based methods (SAM2 Tiny, SAM3) degrade after R1. Box baselines (markers at R0) cannot iterate.}
    \label{fig:refinement_curve}
\end{figure}

\begin{figure}[H]
    \centering
    \includegraphics[width=\linewidth]{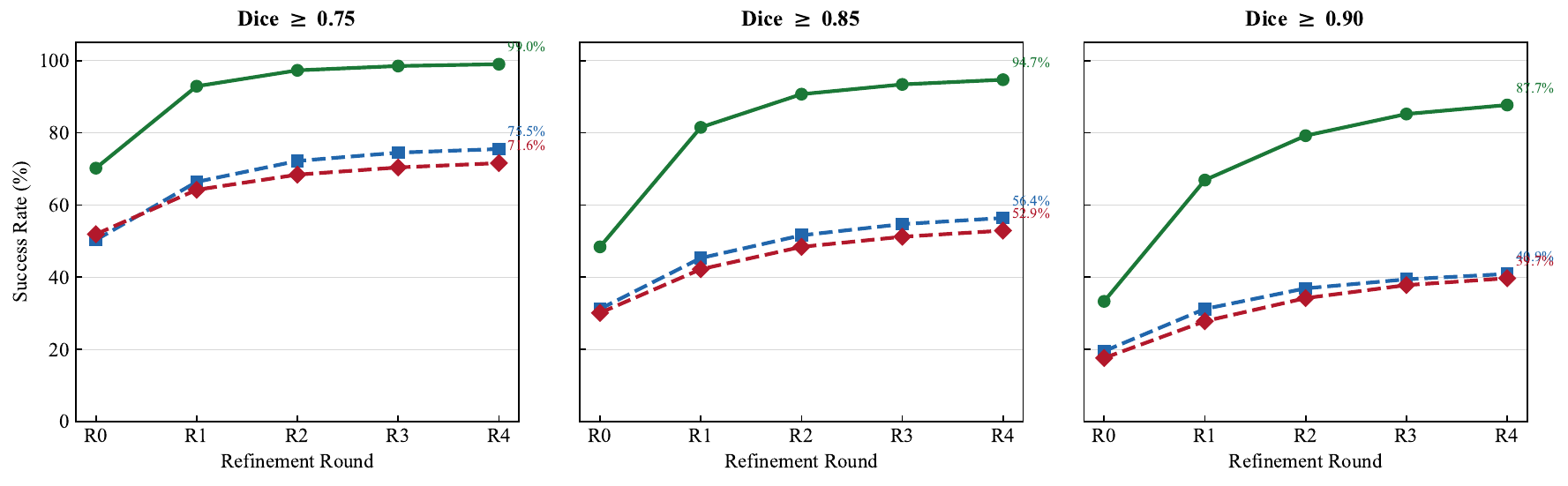}\\[2pt]
    \includegraphics[width=0.45\linewidth]{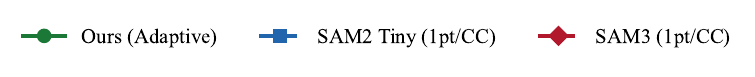}
    \caption{Cumulative success rate on \dataseta{} at three Dice thresholds. Our method converges to high-quality masks within 1--2 rounds for most samples, while point-based methods plateau early.}
    \label{fig:success_rate}
\end{figure}

\begin{figure}[H]
    \centering
    \includegraphics[width=0.85\linewidth]{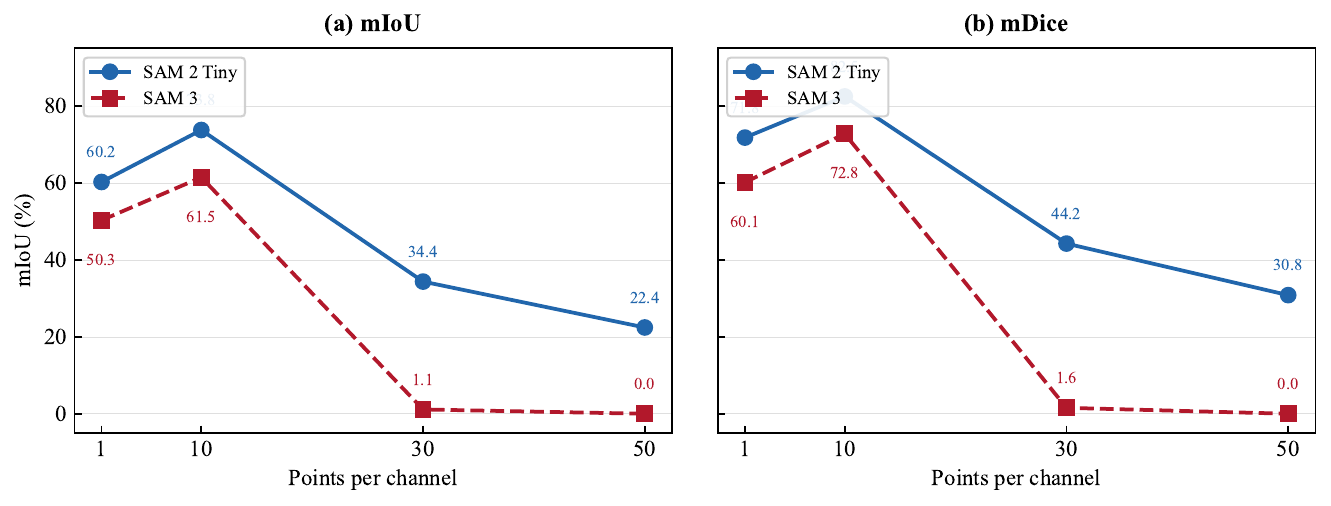}\\[2pt]
    \includegraphics[width=0.3\linewidth]{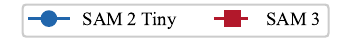}
    \caption{Point prompt degradation on \datasetb{}. Both \samtwo{} Tiny and \samthree{} peak at 10 points and collapse beyond 30, confirming that the point interface cannot exploit denser spatial signals.}
    \label{fig:point_degradation}
\end{figure}

\begin{figure}[H]
    \centering
    \includegraphics[width=0.85\linewidth]{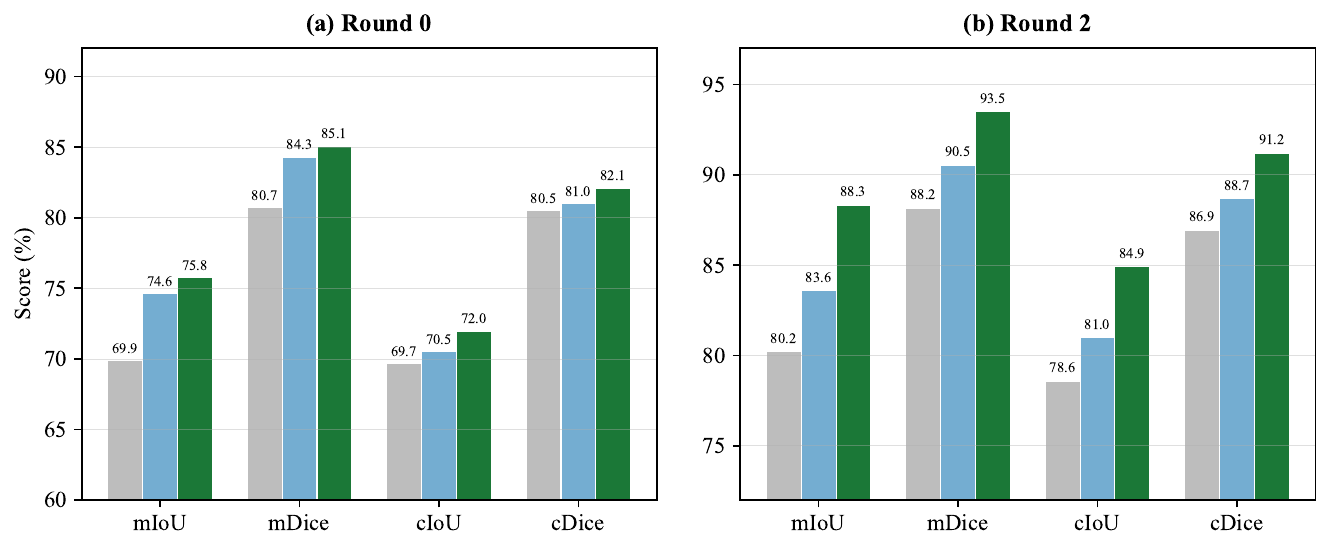}\\[2pt]
    \includegraphics[width=0.4\linewidth]{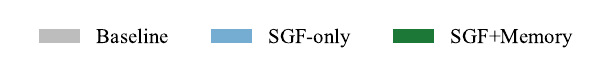}
    \caption{Component ablation on \dataseta{}. Each module (SGF, Memory) contributes additive gains at both R0 and R2 across all four metrics. Numeric values are in Table~\ref{tab:component_ablation}.}
    \label{fig:component_ablation}
\end{figure}

We include supplementary visualizations that complement the main-text analysis. Fig.~\ref{fig:refinement_curve} plots the iterative refinement curves corresponding to Table~\ref{tab:main_allrounds}. Fig.~\ref{fig:success_rate} shows cumulative success rates at three Dice thresholds. Fig.~\ref{fig:point_degradation} illustrates the point prompt degradation discussed in Sec.~\ref{sec:eval_protocol}. Fig.~\ref{fig:component_ablation} visualizes the component ablation from Table~\ref{tab:component_ablation}.

\bibliographystyle{splncs04}
\bibliography{references}

\end{document}